\documentclass[prl,twocolumn, aps,showpacs,superscriptaddress]{revtex4}

\usepackage{latexsym}
\usepackage{amsmath}
\usepackage{amssymb}
\usepackage{amsfonts}
\usepackage{ifthen}
\usepackage{revsymb}
\usepackage{graphicx}

\usepackage{graphicx}
\usepackage{amsmath}
\usepackage{amssymb}
\usepackage{amsthm}

\newcommand{\be}{\begin{equation}}
\newcommand{\ee}{\end{equation}}
\newcommand{\ba}{\begin{array}}
\newcommand{\ea}{\end{array}}
\newcommand{\bea}{\begin{eqnarray}}
\newcommand{\eea}{\end{eqnarray}}

\newcommand{\ket}[1]{| #1 \rangle}
\newcommand{\bra}[1]{\langle #1 |}

\begin{document}

\title{From Majorana Fermions to Topological Order}

\author{Barbara M. \surname{Terhal}}
\author{Fabian  \surname{Hassler}}
\affiliation{Institute for Quantum Information, RWTH Aachen University, 52056 Aachen, Germany}
\author{David P. \surname{DiVincenzo}}
\affiliation{Institute for Quantum Information, RWTH Aachen
University, 52056 Aachen, Germany}
\affiliation{Dept. Theoretical Nanoelectronics, PGI, Forschungszentrum Juelich, 52425 Juelich, Germany}

\date{\today}

\begin{abstract}
We consider a system consisting of a 2D network of links between
Majorana fermions on superconducting islands. We show that the fermionic
Hamiltonian modeling this system gives rise to Kitaev's toric code in
fourth-order perturbation theory. By using a Jordan-Wigner transformation we can map the model onto a family of signed 2D Ising models in a transverse field where the signs, ferromagnetic (FM) or anti-ferromagnetic (AFM), are determined by additional gauge bits. Our mapping allows an understanding of the non-perturbative regime and the phase transition to a non-topological phase. We discuss the physics behind a possible implementation of this model and argue how it can be used for topological quantum computation.
\end{abstract}


\maketitle

Kitaev's well-known toric code \cite{kitaev:top,wen:model} is a toy model Hamiltonian which demonstrates the concept of topological order in two dimensions. One can imagine storing a qubit in this ground space of this model such that, at low temperature $T$ compared to the gap, dephasing of such qubit is exponentially suppressed with growing lattice size.
More general constructions allow the encoding of many qubits in the ground space and the topological implementation of a cNOT and Hadamard gate by means of Hamiltonian or code deformation \cite{dennis+:top, RHG:threshold, bombin_martindelgado, fowler+:unisurf}. 
\begin{center}
\begin{figure}[htb]
\centerline{
\mbox{
\includegraphics[width=0.9\linewidth]{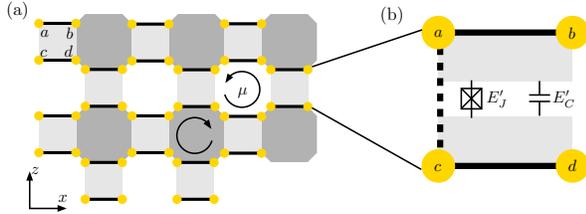}}}
\caption{(Color online) (a) Fermionic model studied in this paper. Each
island (light gray square) has four Majorana fermions (yellow dots)
labeled as $a,b,c,d$. $50\%$ occupancy is favored for these two 
fermionic modes, as expressed by the parity constraint in Eq.~(\ref{eq:h0}). A
weaker quadratic interaction exists between Majorana fermions on
islands $i$ and $j$ along diagonal links $V_{i=\mu\pm \hat{z},j=\mu
\pm \hat{x}}$, Eq.~(\ref{def:links}). (b) Zoomed view of a single island.  The Majorana wire has a C shape (black line)
in order to be able to tune the overlap between the $a$- and $c$-Majorana and
thus implement an $X$-gate. A $Z$-gate/measurement is implemented by 
increasing the ratio of $E_C'/E_J'$, see main text.}
 \label{fig:model}
\end{figure}
\end{center}

In this paper, we investigate how one could arrive at the toric code Hamiltonian starting from a
realistic model of interacting fermions. Kitaev has shown how the
toric code emerges in fourth-order perturbation theory from the
so-called honeycomb model \cite{kitaev:anyon_pert} (see also \cite{brell+:gadgets}). Given
the recent interest in making proximity-coupled semiconducting nanowires
which support weakly-interacting Majorana bound states at their ends
\cite{kitaev:01,fu:08,lutchyn:10,oreg:10,mourik:12,deng:12,rokhinson:12}, we believe that our model may
provide a viable route to the realization of topological quantum computation. The interest in
Majorana fermion wires is partially motivated by their fermionic-parity
protected ground space degeneracy which allows parity protected quantum
computation \cite{hassler:10,sau:10,beenakker:11} and braiding in
networks of nanowires \cite{alicea:11,sau:11,heck:11,halperin:11}. 
The advantage of the approach advocated in this paper is that the protection
is fully topological and no longer based on fermionic-parity conservation.
The idea of engineering a topologically-ordered Hamiltonian using Josephson-junction arrays
has been explored mostly in the work of Ioffe {\em et al.}, see e.g. \cite{gladchenko}.

We consider the following fermionic Hamiltonian $H=H_0+V$ where
$H_0=\sum_{i}^{2L^2} H_0^i$ and $i$ labels the square islands in
Fig.~\ref{fig:model}. The lattice in
Fig.~\ref{fig:model}(a) has periodic boundary conditions in both directions (see the Appendix for a discussion of the model with open boundaries). Each $H_0^i$ acts on two fermionic modes or four Majorana modes as
\be
H_0^i=-\Delta c_a^i c_b^i c_c^i c_d^i.
\label{eq:h0}
\ee
Further, we have $V=\lambda \sum_{i < j} V_{i,j}$ (with $\lambda > 0$) where $i,j$ represents the
interaction between two Majorana fermions on adjacent islands $i$ and $j$,
i.e., $V_{i=\mu\pm \hat{z},j=\mu\pm \hat{x}}$ for a plaquette $\mu$ equals
\begin{eqnarray}
V_{\mu+\hat{z},\mu-\hat{x}}=\pm i c_b^{\mu-\hat{x}} c_c^{\mu+\hat{z}}, &  V_{\mu-\hat{z},\mu-\hat{x}}=\pm i c_a^{\mu-\hat{z}} c_d^{\mu-\hat{x}}, \nonumber \\
V_{\mu-\hat{z},\mu+\hat{x}}=\pm i c_c^{\mu+\hat{x}} c_b^{\mu-\hat{z}}, & V_{\mu+\hat{z},\mu+\hat{x}}=\pm i  c_d^{\mu+\hat{z}} c_a^{\mu+\hat{x}}.
\label{def:links}
\end{eqnarray}
\begin{center}
\begin{figure}[tb]
\centerline{
\mbox{
\includegraphics[height=4cm]{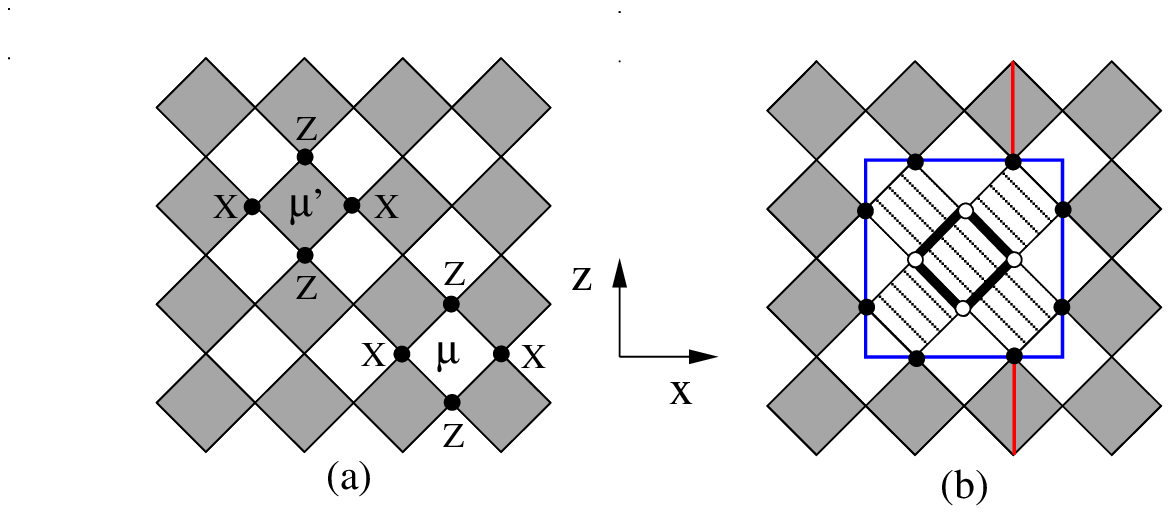}}}
\caption{(Color online) (a) Toric code on a $L \times L$ lattice with $2L^2$ qubits on vertices. The Hamiltonian is a sum over white and grey plaquette operators $A_{\mu}=Z_{\mu+\hat{z}}Z_{\mu-\hat{z}}X_{\mu+\hat{x}}X_{\mu-\hat{x}}$. (b) When the hatched plaquettes are omitted in the toric code Hamiltonian, the ground space degeneracy increases from 4 to 8. The logical operators of this additional qubit are depicted by the blue and red lines. We can call this a white hole qubit as it is obtained by making a hole in the lattice which is centered around a white plaquette.}
 \label{fig:toric}
\end{figure}
\end{center}

All link operators $V_{\mu\pm \hat{z},\mu\pm \hat{x}}$ mutually commute. The
$\pm$ signs of these terms will be fixed according to the consistent orientation of the 
plaquettes in Fig.~\ref{fig:model}, i.e., the link on the top-left of a
white plaquette $\mu$ represents the interaction $V_{\mu+\hat{z},\mu-\hat{x}}=i
c_b^{\mu-\hat{x}} c_c^{\mu+\hat{z}}$.  Physically, the signs depend on
microscopic detail and thus we assume them to be random but fixed. We can find an extensive set of operators which commute with all terms of $H$ and which all mutually commute. These are, first of all, weight-8 fermionic plaquette operators $\{C^g_{\mu},C^{w}_{\mu}\}$ (where $g$ ($w$) stands for gray (white) plaquettes $\mu$) which are the product of four link operators around a plaquette:
\be
C^{g/w}_{\mu}= c_d^{\mu+\hat{z}} c_a^{\mu+\hat{x}} c_c^{\mu+\hat{x}}c_b^{\mu-\hat{z}} c_a^{\mu-\hat{z}} c_d^{\mu-\hat{x}} c_b^{\mu-\hat{x}}c_c^{\mu+\hat{z}}.
\label{def:loops}
\ee

Secondly, the torus has two homologically non-trivial closed loops $\gamma_1,\gamma_2$ and the loop operators $C_{\gamma_{1/2}}=\Pi_{(i,j) \in \gamma_{1/2}}V_{i,j}$ commute with all link $V_{i,j}$ and island operators $H_0^i$. Thus the Hamiltonian is block-diagonal with respect to subspaces (`sectors') characterized by the eigenvalues $C^{g/w}_{\mu}=\pm 1$ and $C_{\gamma_{1/2}}=\pm 1$. 

We analyze the model in the perturbative regime where $\Delta \gg \lambda$, see e.g. \cite{kitaev:anyon_pert}. The ground space of $H$ at $\lambda=0$ is characterized by $\{c_a^i c_b^i c_c^i c_d^i=+1\}$, and thus the ground space on each island is a two-dimensional subspace, a qubit. One can define the logical $X$ and $Z$ operator on this island qubit as 
\begin{eqnarray}
X_i=i c_c^i c_a^i \equiv i c_d^i c_b^i, & Z_i=i c_c^i c_d^i \equiv i c_b^i c_a^i.
\label{eq:pauli}
\end{eqnarray}
Let $P_-=2^{-2L^2}\Pi_{i=1}^{2L^2}(I+c_a^i c_b^i c_c^i c_d^i)$ be the projector onto this $2^{2L^2}$-dimensional unperturbed ground space. The first non-trivial term in the perturbative expansion occurs in $4^{th}$ order (see the Appendix), i.e.
\begin{equation}
H_{\rm eff}=-\frac{5\lambda^4 }{16\Delta^3}\sum_{\mu=1}^{2L^2} A_{\mu}+O\left(\frac{\lambda^6}{\Delta^5}\right).
\label{eq:pert}
\end{equation} 
where $A_{\mu}=Z_{\mu+\hat{z}} X_{\mu+\hat{x}} Z_{\mu-\hat{z}} X_{\mu-\hat{x}}$, i.e., the plaquette terms of the toric code in Fig.~\ref{fig:toric} \cite{familiar}. Note that  $P_- C^{g/w}_{\mu} P_-=A^{g/w}_{\mu}$ and hence the four-dimensional toric code ground space of $H$ when $\Delta \gg \lambda$ lies in the $\{C^{g/w}_{\mu}=+1\}$ sector. 
\begin{center}
\begin{figure}[tb]
\centerline{
\mbox{
\includegraphics[height=2cm]{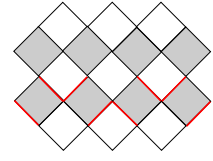}
 }}
\caption{(Color online) The gauge bits $\sigma$ set the Ising interactions to FM (black edges) except for an AFM (red edges) loop around the torus. This AFM boundary will be felt in the ferromagnetic (FM) phase, but not in the paramagnetic (PM) phase of the model, leading to degeneracy. A loop operator $C_{\gamma}$ in the fermionic model becomes a product of Ising edges which winds around the torus.}
 \label{fig:loop}
\end{figure}
\end{center}
Let us consider how the topological phase extends to the regime
where $\frac{\Delta}{\lambda}=O(1)$. We can get insight for this regime by considering higher-order terms in the
perturbative expansion (see e.g. \cite{vsd:honey} for such expansion for the honeycomb model), see the Appendix. This calculation shows that the ground space degeneracy of the toric code is broken only in $(2L)^\text{th}$-order perturbation theory and we expect that the topological phase will destabilize via this mechanism.

\begin{center}
\begin{figure}[tb]
\centerline{
\mbox{
\includegraphics[width=0.7\linewidth]{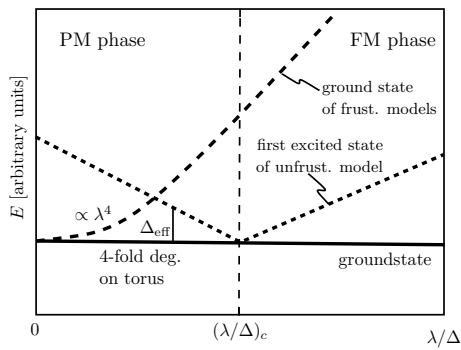}
 }}
\caption{Sketch of the spectrum of system as a function of $\lambda/\Delta$. For
  small $\lambda$, the system is in a topological state with a four-fold
  ground state degeneracy on the torus. The first excited states for small $\lambda$
  are Ising models with frustration as determined by the gauge bits. All these
  models are degenerate for $\lambda=0$; and the degeneracy lifts in forth
  order perturbation theory in $\lambda$, see Eq.~\eqref{eq:pert}. The phase transition 
to a state without topological order happens
  at the transition point $(\lambda/\Delta)_c$ of the unfrustrated Ising
  model. At this transition the gap of the Ising
  model closes and the degeneracy of the topological states vanishes.} 
 \label{fig:spectrum}
\end{figure}
\end{center}

To consolidate this picture we map our model via a Jordan-Wigner (JW)
transformation, onto a family of signed transverse field (TF) Ising models on a 2D square lattice, see Appendix. The JW mapping preserves the locality of the interactions, and all reductions are explicit and rigorous. Previous work \cite{FSS} has shown how to map an Ising gauge theory in a transverse field (assuming open boundary conditions) by a Jordan-Wigner transformation onto the Majorana fermion model discussed here.
Our mapping clarifies the nature of the topological phase transition and the parameter values for which it will occur.  In this analysis we restrict ourselves to the sector $\{C^w_{\mu}=+1\}$. Note that when $\Delta=0$ the ground state of $H$ is unique since we are in a state of fixed link parity $\{V_{i,j}=-1\}$. This state lies in the $\{C^{g/w}_{\mu}=+1\}$ sector since each $C^{g/w}_{\mu}$ is a product of four link operators. As we saw above, at $\lambda \ll \Delta$, the ground space also lies in the $\{C^{g/w}_{\mu}=+1\}$ sector. 
The mapping decouples our fermionic model into a set of transverse field Ising models $H(\sigma)=-\lambda \sum_{i,j} \sigma_{i,j} S_i^z S_j^z-\Delta \sum_i S_i^x$ with the condition $\Pi_{(i,j) \in \mbox{ \tiny white } \mu} \sigma_{i,j}=1$ so that the sign of the Ising interactions is determined by gauge bits $\sigma_{i,j} \in \{-1,1\}$ associated with the edges. The gauge condition expresses the fact that the white plaquettes are never frustrated, i.e. $C^w_{\mu}=+1$, but a gray plaquette is frustrated when $C^g_{\mu}=\Pi_{(i,j) \in \mbox{\tiny gray } \mu} \sigma_{i,j}=-1$. The spectrum of $H(\sigma)$ solely depends on the frustration of the Ising interactions and the presence of domain walls or homologically non-trivial loops, see Fig.~\ref{fig:loop}. We anticipate the following spectrum, see Fig.~\ref{fig:spectrum}. At both ends of the parameter region ($\lambda$ or $\Delta=0$), the groundspace lies in the unfrustrated TF Ising model sector. We have numerically confirmed this for small lattice sizes for the entire parameter regime. Because of the symmetry between gray and white plaquettes, this finding also motivates the choice for $\{C^w_{\mu}=+1\}$ as the ground sector.  Fig.~\ref{fig:loop} depicts a configuration $\sigma$ which represents a homologically non-trivial loop; all plaquettes are unfrustrated, but an Ising model with such an AFM sign pattern will contain a domain wall of length at least $L$ where bonds are not satisfied. The topological phase is identified as the paramagnetic (PM) phase $\langle S^z \rangle = 0$ in the TF Ising models. In this phase the ground spaces of the Hamiltonians $H(\sigma)$ with unfrustrated configurations $\sigma$ with or without the 2 non-trivial AFM loops are approximately degenerate: this is the topological degeneracy whose splitting we expect to scale as $\exp(-L/\xi)$ where $\xi$ is the correlation length of the TF ferromagnetic Ising model. We expect the effective gap $\Delta_{\rm eff}$ above the degenerate ground-space to monotonically increase before we reach the second-order
phase transition of the TF ferromagnetic Ising model, which is known to occur
around $\left(\frac{\lambda}{\Delta}\right)_c \approx 0.33$ \cite{book:ising}.
Elementary excitations of the toric code with $A_{\mu}=+1$ for two gray plaquettes $A_{\mu}$ correspond to ground-states of TF Ising models with frustration at those particular gray squares.

We can consider the effect of additional perturbations. One would expect quadratic Majorana fermion
perturbations of strength $\epsilon$ on each island due the coupling of the
wire ends on an island. If a qubit is encoded in a pair of Majorana wires, such coupling leads to
an energy-level splitting of the qubit state of strength $\epsilon$. Here the advantage of our topological encoding becomes clear. Consider a perturbation $U \propto \epsilon \,i c_b^{\mu-\hat{x}} c_d^{\mu-\hat{x}}$
where $\mu$ is some white plaquette (by symmetry other perturbations would act
similarly) and $\epsilon \ll \lambda$. On the groundspace of $H_0$, the perturbation acts as a local term $\propto \epsilon X_{\mu-\hat{x}}$, hence we expect that the topological degeneracy is preserved up to some critical strength $(\frac{\epsilon}{\Delta_{\rm eff}})_c$ where $\Delta_{\rm eff}$ is the gap above the degenerate ground-space in the topological phase. In practice, we expect these quadratic perturbations to be random (and weak) and hence they could be beneficial in stabilizing the topological quantum memory at finite temperature by limiting the diffusion of anyons (see e.g. \cite{starketal}).

We analyze the possibility of implementing the model presented above in
superconducting-semiconducting heterostructures. Putting a semiconducting
nanowire such as InAs or InSb with strong spin-orbit interaction on top of a
conventional superconductor subject to a sufficiently strong magnetic field
leads to a pair of Majorana modes located at the ends of the nanowire
\cite{lutchyn:10,oreg:10}. We focus on a 2D array of superconducting
islands each supporting two nanowires leading to four unpaired Majorana
modes, see Fig.~\ref{fig:model}. The product of the four Majorana operators
$\mathcal{P}^i=c^i_ac^i_bc^i_cc^i_d $ is fixed by the \emph{parity}
$\mathcal{P}^i = (-1)^{n_i}$ of the number of electrons $n_i$ on the
$i$-th island \cite{fu:10,xu_fu:fracmodel}. Two Majorana modes $c^i_x$
and $c^j_y$ on different superconducting islands $i$ and $j$ interact with
each other via the anomalous Josephson interaction
\begin{equation}\label{eq:josephson}
  H_J = \sum_{i,j} \Gamma_{i,j} V_{i,j} \cos [ (\phi_j -\phi_i)/2],
\end{equation}
where $\phi_i$ denotes the superconducting phase on the $i$-th island and
$\Gamma_{i,j}$ is proportional to the probability amplitude for tunneling
a single electron/hole across the link $i,j$ from mode $c^i_x$ to $c^j_y$
between the islands $i$ and $j$ \cite{kitaev:01}. Along the lines of
Ref.~\cite{heck:11}, we shunt each superconducting island with a strong
Josephson coupling $E_{J}$ to a common ground superconductor.  This Josephson
coupling fixes the superconducting phases $\phi_i$ to
a common value due to large fluctuations of the number of electrons (in
units of two) on and off the island. Note that this way of freezing out the
bosonic degrees of freedom due to the superconductivity is different from
the one discussed in Ref.~\cite{xu_fu:fracmodel} which proposes a
large charging energy which fixes the number of electrons by delocalizing
the superconducting phase completely.  Even though the charge is strongly
fluctuating, the fermion parity $\mathcal{P}^i$ remains conserved. In the
ideal case when all the Josephson couplings are equal $\Gamma_{i,j} =
\lambda$, the anomalous Josephson interaction $H_J$ implements $V$. Of course,
$V$ in general will not have the orientation indicated in Fig.~\ref{fig:model}, but if we work 
with a lattice with open boundary conditions (corresponding to the surface code \cite{BK:surface}), any sign pattern
of the link interactions simply picks out a topological sector with a corresponding pattern of eigenvalues 
$A_{\mu}=\pm 1$ as the ground-space \cite{toric}, see our discussion in the Appendix. From the coding perspective it is well known that topological computation can proceed not just in the trivial syndrome (all eigenvalues of $A_{\mu}=+1$) sector but also in some non-trivial syndrome sector.

Next, we discuss the effect of a capacitive coupling
\begin{equation}\label{eq:capacitance}
H_C = E_C \sum_i (n_i - n_\text{ind})^2
\end{equation}
to the ground plate with the capacitive energy $E_C = e^2/2C$ where $C$
is the capacitance of the island with respect to ground and $n_\text{ind}$
a constant which is due to gate voltages. For simplicity of notation, we
have assumed all the capacitances $C$ and offset charges $n_\text{ind}$ to
be equal. In the regime with $\lambda \lesssim E_C \ll E_J$, the capacitive
coupling introduces phase-slips through the strong Josephson junctions
and thus an energy difference between states with different fermion
parities. This leads to Eq.~\eqref{eq:h0} with $\Delta \propto E_C^{1/4} E_J^{3/4}\cos (\pi
n_\text{ind})e^{-\sqrt{8 E_J/E_C}}$ \cite{heck:11}. The sign of $\Delta$
depends on the value of $n_\text{ind}$ can thus be tuned in principle.
A residual interaction between two Majorana modes $c^i_x$ and $c^i_y$
with strength $\epsilon$ is due to overlap of the wave functions of the
Majorana bound states. However, because the states are localized, this
coupling can be made exponentially small by keeping the modes sufficiently
far apart from each other.  Having sufficiently strong tunneling coupling
between the Majorana fermions along the links, a value $\lambda \simeq
200\,$mK seem realistic as the bare gap of the Majorana wire is likely of
the order of a few K \cite{alicea:11}. Choosing the Josephson energy to be
$E_J \simeq 10\,$K and a capacitive energy $E_C \simeq 5\,$K, we obtain
$E_J \gg \Delta \gtrsim \lambda$ as required . The resulting optimal value
of $\Delta_\text{eff}$ will be of the order of $\lambda \simeq
200\,$mK.

It is possible to tune the tunnel coupling $\Gamma_{i,j}$ by changing the tunneling barrier between island $i$ and $j$
by a nearby gate. This is a mechanism through which we can create holes 
in the lattice. In practice one can work with a lattice of superconducting islands which represents
the surface code with open boundary conditions, encoding one logical qubit. 
One can apply the ideas of the surface code architecture if (\emph{i}) one can
make (and move) gray and white holes of arbitrary size in this surface and (\emph{ii}) one can locally measure $X_i$ and $Z_i$ and prepare $X_i$ and $Z_i$ eigenstates. The preparation of certain 1-qubit ancillas can then be used to achieve universality \cite{bk:magicdistill}. An example of a white hole qubit is depicted in Fig.~\ref{fig:toric}(b). Such a white hole can be obtained by cutting off the four black links surrrounding the center white 
plaquette such that the hatched plaquette terms no longer appear in the effective Hamiltonian. Moving such a white hole
could be done by adiabatically changing the strength of Majorana links in order to turn links on and off. The operations (\emph{ii}) can be implemented using the setup of
Fig.~\ref{fig:model}(b).  Instead of a single superconducting island, each
site consist in fact of two islands with two Majorana modes each.  Most of the
time, these islands are coupled to each other via a strong Josephson coupling
$E_J' \gg E_C',E_C$ (see Fig.~\ref{fig:model}) such that they essentially act as a single island such
that all the discussion above applies unchanged.  Increasing the ratio
$E_C'/E_J'$ turns on a magnetic field along the $z$-axis which can be used to
implement rotations around this axis.  Additionally, the measurement of $Z_i$
can be implemented by coupling one of the superconducting islands to a fermion
parity meter \cite{hassler:10}.  Single qubit universality is achieved by
increasing the overlap of the $a$ and $c$ Majorana modes---by decreasing the
length of the topological trivial part of the Majorana wire indicated by the
dashed---and thus effectively implementing a magnetic field along the $x$-axis.

We acknowledge fruitful discussions with A. Akhmerov. DDV and FH are grateful for support from the Alexander von Humboldt foundation.

\section*{Appendix}

We start by detailing some of the steps in the perturbative analysis. We defined $P_-=2^{-2L^2}\Pi_{i=1}^{2L^2}(I+c_a^i c_b^i c_c^i c_d^i)$ as 
the projector onto the $2^{2L^2}$-dimensional groundspace and $P_-+P_+=I$. Let $V_{\pm \mp}=P_{\pm} V P_{\mp}$ and let 
$G_{+}=P_+ H_0^{-1} P_+$ where we have redefined $H_0$ as $H_0+2\Delta L^2$ such that its lowest-eigenvalue is 0. 
 In the self-energy expansion all terms with odd number of perturbations $V$ vanish. The second-order
term $V_{-+} G_+ V_{+-}$ contributes a term proportional to $I$ whereas the fourth-order term equals
\be\label{eq:pert_detail}
V_{-+} G_+ V_{++} G_+ V_{++} G_+ V_{+-}=-\frac{5\lambda^4 }{16\Delta^3}\sum_{\mu=1}^{2L^2} A_{\mu},
\ee
where $A_{\mu}=Z_{\mu+\hat{z}} X_{\mu+\hat{x}} Z_{\mu-\hat{z}} X_{\mu-\hat{x}}$, i.e., the plaquette terms of the toric code in
Fig.~\ref{fig:toric} \cite{familiar}. Note the difference with Kitaev's honeycomb model where the prefactor of these plaquette terms is $\frac{-1}{16}$ instead of $\frac{-5}{16}$; this
is because the links in our model commute whereas the links in Kitaev's model mutually anti-commute on their common qubit. All higher-order terms in the perturbative expansion 
consist of loops of links on the lattice and can be represented as products of plaquette operators $A_{\mu}$, i.e.
the terms are of the form $-\sum_{\mu_1}\ldots \sum_{\mu_k} A_{\mu_1}\ldots A_{\mu_k}$, all {\em with the same negative sign}, further stabilizing the ground space with $\{A_{\mu}=+1\}$.

In $(2L)^\text{th}$-order one obtains also terms proportional to $A_{\gamma} \equiv P_0 C_{\gamma}
P_0 \propto \Pi_{i \in \gamma}\sigma_i$  where $i$ is a product over islands
through which the loop $\gamma$ goes and $\sigma_i=X_i$, $Y_i$ or $Z_i$ as
defined in Eq.~(\ref{eq:pauli}) depending on the loop $\gamma$ taking one of the three directions across island $i$. Since $A_{\gamma_1}$ and $A_{\gamma_2}$ commute with all $\{A_{\mu}\}$ and with each other, they must leave the ground space of the toric code invariant and be mutually commuting 
products of the 4 logical operators of the toric code qubits, i.e. they can be represented as $\overline{X}_1 \overline{X}_2$ and $\overline{Z}_1 \overline{Z}_2$ where 
$(\overline{X}_i,\overline{Z}_i)$, $i=1,2$ are the nonlocal logical operators of the 2 toric code qubits. Hence the presence of these terms in the effective Hamiltonian breaks the topological degeneracy.

\begin{center}
\begin{figure}[tb]
\centerline{
\mbox{
\includegraphics[height=4cm]{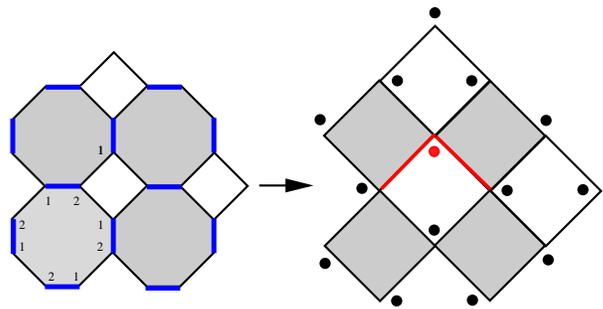}}}
\caption{(Color online) The blue horizontal and vertical links denote $ZZ$ terms in the Hamiltonian and the black diagonal links are proportional to $XX$ on the square-octagon lattice. By laying out a basis of Bell states on the blue links, we can map the Hamiltonian on the square-octagon lattice to a family of signed Ising Hamiltonians in a transverse field on a square lattice. The positions of qubit $1$ and $2$ of each Bell state is indicated so that $XX$ links always act between a qubit $1$ and $2$ of different Bell states and we restrict ourselves to even $L$. In the unitarily transformed model, every Ising qubit has an additional sign qubit (black dots) next to it. If the sign qubit is $\ket{1}$ (marked as red), it flips the adjacent red edges from ferromagnetic to anti-ferromagnetic. We can alternatively associate a gauge bit $\sigma_{i,j}$ with every edge $(i,j)$ if we ensure that the product of $\sigma_{i,j}$ around a white plaquette is $1$.}
 \label{fig:square_oct}
\end{figure}
\end{center}

The overall mapping via the JW transformation goes as follows.  We will denote the eigenvalues of the operators $C^w_{\mu}$, Eq. (\ref{def:loops}), as $c^w_{\mu}$. Note, by the way, that there is one linear dependency between the plaquette operators  i.e. $C_{\rm all}=\Pi_{\mu} C^g_{\mu}=\Pi_{\nu}C^w_{\nu}$ where $C_{\rm all}$ is the product of all $8L^2$ Majorana operators. First, we map our model via a JW
transformation onto a model which for {\em fixed} eigenvalues $\{c_\mu^w\}$
consists of $XX$ (strength $\lambda$) and $ZZ$ links (strength $\Delta$), a
square-octagon model on the left in Fig.~\ref{fig:square_oct}. We consider this
square-octagon model for $\{c_{\mu}^w=+1\}$, but we extend the state space to any state with $\Pi_\mu
C^w_{\mu}=+1$ \cite{proc}. In this extended state space, we use the additional symmetry of the square-octagon model to lay out a basis of Bell states on the islands. In the Bell basis, the square-octagon model reduces to a set of transverse field Ising models where the sign of the Ising interactions is determined by an additional sign qubit. In principle this degree of freedom is present at every island; the sign qubits are the black and red dots in Fig.~\ref{fig:square_oct}. However, these models are unitarily equivalent to ones in which associate a gauge bit $\sigma_{i,j}\in \{-1,1\}$ with every edge $(i,j)$ and fix the gauge $\Pi_{(i,j) \in \mbox{\tiny white } \mu} \sigma_{i,j}=1$. Thus we obtain the transverse field Ising models are $H(\sigma)=-\lambda \sum_{i,j} \sigma_{i,j} S_i^z S_j^z-\Delta \sum_i S_i^x$ where $\Pi_{(i,j) \in \mbox{ \tiny white } \mu} \sigma_{i,j}=1$.  Let us now explicitly show these steps. 

\begin{center}
\begin{figure}[tb]
\centerline{
\mbox{
\includegraphics[height=4cm]{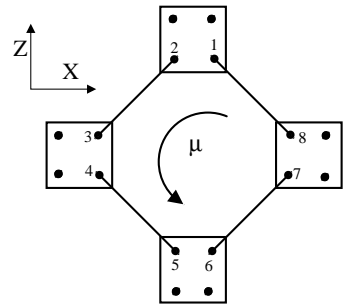}
 }}
\caption{White plaquette where the anti-clockwise direction fixes the sign of the interaction $V$ depicted by black links. For such a white plaquette we choose the Majorana fermion ordering indicated by the numbers 1,2...8 to do the Jordan-Wigner transformation.}
 \label{fig:3}
\end{figure}
\end{center}

For the JW transformation we choose an order for the $8L^2$ Majorana fermions $c_1,\ldots c_{8L^2}$. We
will order the Majorana fermions around every white plaquette as in
Fig.~\ref{fig:3}, i.e., we start with an arbitrary white plaquette, choose
this ordering of labels and go onto any next white plaquette and continue
until we are done labeling all Majorana fermions. Given this ordering the JW transformation maps $\Upsilon(c_{2i-1})=Z_{1}\ldots Z_{i-1}X_i$
and $\Upsilon(c_{2i})=Z_1 \ldots Z_{i-1} Y_i$ where the Pauli's here are of
course unrelated to the Pauli's defined in Eq.~(\ref{eq:pauli}). Under this JW transformation the links $V_{i=\mu\pm \hat{z},j=\mu\pm \hat{x}}$ around a white plaquette $\mu$ become
\begin{eqnarray}
V_{\mu+\hat{z},\mu-\hat{x}}\rightarrow X_{\mu+\hat{z}} X_{\mu-\hat{x}}, & \!\!V_{\mu-\hat{z},\mu-\hat{x}}\rightarrow X_{\mu-\hat{x}} X_{\mu-\hat{z}}, \nonumber \\
V_{\mu-\hat{z},\mu+\hat{x}}\rightarrow X_{\mu-\hat{z}} X_{\mu+\hat{x}}, & \!\!V_{\mu+\hat{z},\mu+\hat{x}}\rightarrow Y_{\mu+\hat{z}} Z_{\mu-\hat{x}} Z_{\mu-\hat{z}} Y_{\mu+\hat{x}}. \nonumber
\end{eqnarray}
The island operator of an island $i$ which is north or south of a white plaquette becomes $H_0^i \rightarrow -\Delta Z_{i,1} Z_{i,2}$ where $i,1$ and $i,2$ label the two qubits on the island, see the blue links in Fig.~\ref{fig:square_oct}. 
The island operator of an island $i$ which is east or west of a white plaquette becomes $H_0^i \rightarrow +\Delta Z_{i,1} Z_{i,2}$. In this qubit representation the white plaquette operator equals $C_{\mu}^{w} \rightarrow -\Pi_{i \in \partial \mu} Z_i$ where $\mu$ now represents a square in Fig.~\ref{fig:square_oct} and $i\in \partial \mu$ is the product over qubits at the corners of the square. Let the subspace with fixed eigenvalues $\{c^w_{\mu}\}$ be denoted as ${\cal H}(\{c^w_{\mu}\})$. We can represent the action of the weight-4 link operator on ${\cal H}(\{c^w_{\mu}\})$ as
\begin{equation}
V_{\mu+\hat{x},\mu+\hat{z}}=Y_{\mu+\hat{z}} Z_{\mu-\hat{x}} Z_{\mu-\hat{z}} Y_{\mu+\hat{x}}=c^w_{\mu}  X_{\mu+ \hat{z}} X_{\mu+\hat{x}}.
\end{equation}
Note that on the space ${\cal H}(\{c^w_{\mu}\})$ the gray plaquette operator equals $C^g_{\mu}=c^w_{\mu-\hat{z}-\hat{x}} \Pi_{i \in \partial \mu} X_i$. 
This implies that on each subspace ${\cal  H}(\{c^w_{\mu}\})$ the action of the Hamiltonian can be best represented using a square-octagon lattice in which qubits live on the vertices, see Fig.~\ref{fig:square_oct}. The white plaquettes have been transformed into squares and the gray plaquettes into octagons. The islands have been stretched in the horizontal and vertical directions and are represented by horizontal and vertical links equal to $\pm\Delta Z_{i,1} Z_{i,2}$  ($-1$ for vertical links, $+1$ for horizontal links). On this lattice, diagonal links between vertices $i$ and $j$ are all $\lambda X_i X_j$ except the diagonal links $\lambda c^w_{\mu} X_{\mu+\hat{z}} X_{\mu+\hat{x}}$ for the white squares labeled by $\mu$. In the remaining analysis we choose $\{c^w_{\mu}=+1\}$, but all further steps can be done for other $\{c^w_{\mu}\}$ \cite{lieb}.

In order to use the square-octagon model, we extend ${\cal H}(\{c^w_{\mu}=+1\})$ into the (almost) full state space, obeying only the parity condition $\Pi_{\mu} C^w_{\mu}=+1$.  At the end of this section, we show why this procedure only adds degeneracies to the Hamiltonian, not altering the eigenspectrum. This parity constraint $\Pi_{\mu} C^w_{\mu}=C_{\rm all}=+1$ is, after the JW transformation, equivalent to $(-1)^{L^2}\Pi_i Z_i=+1$. A new symmetry then becomes apparent, namely the interactions of the square-octagon model commute with vertical and horizontal link operators
$K_j=X_{j,1} X_{j,2}$. Note that by an even number of local unitary $X$ rotations we can rotate all horizontal and vertical links to be $-\Delta
Z_{j,1} Z_{j,2}$. From now on we take $L$ even for simplicity so that the parity constraint implies that $\Pi_i Z_i=1$. A basis for the state space can
then be constructed by laying out Bell states $\ket{\Psi^{st}}\equiv \ket{s,t}$ on the horizontal and vertical links between qubits. Here $s$
represents the sign qubit or eigenvalue of $XX$, i.e., $\ket{s}$
corresponds to $\frac{1}{\sqrt{2}}(\ket{00}+(-1)^{s}\ket{11})$ and
$\frac{1}{\sqrt{2}}(\ket{01}+(-1)^{s}\ket{10})$. Qubit $t$ represents the
eigenvalue of $ZZ$, i.e., $\ket{t}$ corresponds to $(I_1 \otimes X_2^{t})\frac{1}{\sqrt{2}}(\ket{00}\pm \ket{11})$. This choice of basis and local unitary rotation to sign qubits $s$ and Ising qubits $t$ allows one to write the Hamiltonian on the square-octagon lattice as a transverse field Ising model on the $t$ qubits on a square lattice where the additional sign qubits live at every site, see Fig.~\ref{fig:square_oct}.  

In order to obtain this model we lay out the Bell states such that any $XX$ link in the model acts between qubit 1 of one horizontal (resp. vertical) Bell state and qubit 2 of another vertical (resp. horizontal) Bell state, see Fig.~\ref{fig:square_oct}. This necessitates the constraint that $L$ is even. We then use the fact that for two island Bell states $i$ and $j$
\begin{equation}
X_{i,1} X_{j,2} \ket{s, t}_{i} \ket{s',t'}_{j} \rightarrow X^t_i X^t_j Z^s_i \ket{s,t}_{i} \ket{s',t'}_{j},
\end{equation}
where $X_{i,1}$ acts on the first qubit of Bell state $i$ etc. and $X^t_i$ is
a Pauli $X$ on the $t$ qubit of island $i$. Here $Z^s_i$ is Pauli $Z$ on the
sign qubit of island or vertex $i$. The island term $-\Delta Z_{i,1}Z_{i,2}
\rightarrow -\Delta Z^t_i$, i.e., only acting on the $t$ qubit. Note that due to the layout of the sign qubits, see Fig.~\ref{fig:square_oct}, white plaquettes are never frustrated since every sign qubit flips the sign of two Ising edges of the plaquettes. As every Ising qubit has a sign qubit next to it, it implies that a gray square plaquette is frustrated when $C^g_{\mu}=\Pi_{i \in \partial \mu} Z^s_i=-1$.  The parity constraint now reads $\Pi_i Z^t_i=1$.
The transformed Hamiltonian acts with single $Z$s on the sign qubits, hence for every basis state of the sign qubits one obtains a transverse field Ising
model on the $t$ qubits at the vertices of a square lattice on the torus. It is simpler then to represent the effect of the sign qubits by associating them with the edges of the square lattice. We introduce the gauge bit $\sigma_{i,j}\in \{-1,1\}$ for every edge $(i,j)$ which are constrained such that $\Pi_{(i,j)\in \mbox{\tiny white }\mu} \sigma_{(i,j)}=1$. One can show that any TF Ising model with configuration $\sigma$ satisfying this constraint is unitarily equivalent to a TF model with a sign qubit configuration. Thus, modulo unitary transformations, we obtain $H(\sigma)=-\lambda \sum_{i,j} \sigma_{i,j} S_i^z S_j^z-\Delta \sum_i S_i^x$ where the Ising degrees of freedom are the $t$ qubits and the gauge bits are represented by $\sigma$. The gauge constraint is $\Pi_{(i,j) \in \mbox{ \tiny white } \mu} \sigma_{i,j}=1$ and the parity constraint $\Pi_i S_i^x=1$. Note that the parity constraint makes the ground state in the ferromagnetic phase unique, as we expect at $\Delta=0$.

We can also study our perturbative expansion, Eq.~(\ref{eq:pert_detail}), under this mapping. Each TF Ising model $H(\sigma)$ can be handled separately using nondegenerate perturbation theory with a finite radius of convergence $(\frac{\lambda}{\Delta})_c$. 
Perturbing around the (paramagnetic) ground-state $\ket{++ \ldots +}$ for $\lambda=0$, we obtain an effective classical Ising gauge theory (modulo some energy shifts), i.e., 
\begin{eqnarray}
\lefteqn{H_{\rm eff}=  -\lambda \ket{++ \ldots +}\bra{++ \ldots +}\, \times} & \nonumber \\
& \left(\frac{\lambda^3}{\Delta^3} \sum_{\mu} \Pi_{(i,j) \in \mu} \sigma_{i,j}+\mbox{higher order loops}\right),
\end{eqnarray} 
where $\ket{++\ldots +}$ is the transverse field ground-state. 


Consider what happens in this analysis when the sign of the links, Eq.~(\ref{def:links}), is randomly changed (and fixed) with respect to the orientation in Fig.~\ref{fig:model}. Now we should no longer a priori select $\{c^w_{\mu}=+1\}$ as the ground sector. However, all steps can be carried through as long as we retain the signs $\{c^w_{\mu}\}$. A sign change in the links represent a sign change in the $XX$ links of the family of Hamiltonians on the square-octagon lattice, hence white plaquettes can now be frustrated. The proper choice of eigenvalues $\{c^w_{\mu}=\pm 1\}$ for the ground sector is the one in which no square originating from a white plaquette is frustrated. This choice may however lead to frustration for the gray plaquettes (since $C^g_{\mu} \rightarrow c^w_{\mu-\hat{z}-\hat{x}} \Pi_{(i,j) \in \mbox{\tiny gray }\mu} \sigma_{i,j}$.). If no frustration can be avoided, we may expect the behavior of this model to be intrinsically different; this can happen on the torus if a single link changes sign.  A single link sign flip makes $A_{\mu} \rightarrow -A_{\mu}$ for a white and a grey plaquette neighboring the link in the effective Hamiltonian $H_{\rm eff}$, Eq. (\ref{eq:pert}). But there is no state with lowest possible energy $-\Delta_{\rm eff}L^2$. There will be frustration and there are 10 degenerate toric code ground states with two possible excitations all having the same energy of $(- L^2+1) \frac{\Delta_{\rm eff}}{2}$.  This problem arises since excitations in the toric code produce defects on an even number of white plaquettes and an even number of gray plaquettes, hence a single sign change of a link does not pick out an excited sector as ground sector. For a physical model with open boundary conditions such as the surface code, there are eigenstates with any pattern of eigenvalues $A_{\mu}=\pm 1$, hence it is possible to avoid frustration. 

Note that in such surface code lay-out, our model will have 4 `unlinked' Majorana fermions at the corners of the lattice, let us call them $c_1,c_2, c_3$ and $c_4$. The Hamiltonian of this Majorana fermion surface code commutes with any string of Majorana operators which starts and ends at one of the four corner Majorana fermions. Such strings anticommute when they touch the same corner Majorana fermion forming a pair of logical operators. Hence the spectrum of this Hamiltonian has a double degeneracy throughout the entire parameter range; it is the encoding of the surface code qubit. When $\Delta=0$, the Hamiltonian commutes with the four corner Majorana fermions, hence the degeneracy is 4 in this regime. This 4-dimensional degeneracy is exponentially protected when $\lambda$ is turned on, but one of these Majorana qubits (with logical operators, say, $\overline{X}=c_1$, $\overline{Z}=c_2$ and $\overline{Y}=i c_1 c_2$) is only protected by fermion-parity. At the phase transition, one looses the fermion-parity protected Majorana qubit, but keeps the Majorana surface code qubit.

Let us add one last comment on how to handle quasi-particle tunneling onto the superconductor island which would change the fermion-parity on the island. Such event can be viewed as a leakage error with respect to our encoding, i.e. an error which maps us outside the proper code or ground space. If one can measure the parity operator $H_0^i$ on an island and project the state back onto the parity ground space, we reduce such leakage error to no error or a regular error which can either be tolerated at low density or be error-corrected actively. 

\subsection*{Projecting Back Lemma}
Let the subspace with fixed eigenvalues $\{c^w_{\mu}\}$ be denoted as ${\cal H}(\{c^w_{\mu}\})$ and the projector onto this subspace as $P(\{c^w_{\mu}\})$. Let $H(\{c^w_{\mu}\})$ be the fermionic Hamiltonian after the JW transformation with fixed eigenvalues $\{c^w_{\mu}\}$, i.e the Hamiltonian with links on the square-octagon lattice, Fig.~\ref{fig:square_oct}. We want to consider the spectrum of the Hamiltonian $H(\{c^w_{\mu}\})$ on a larger state space only obeying the parity condition $\pi=\Pi_{\mu} C^w_{\mu}=+1$ and show that the spectrum is identical to that on ${\cal H}(\{c^w_{\mu}\})$ except for additional degeneracies. Let $P_{\pi}$ be the projector onto the space of states with $\pi=\Pi_{\mu} C^w_{\mu}=+1$. We prove that all eigenstates of $H(\{c^w_{\mu}\})$ for which $P_{\pi}\ket{\psi} \neq 0$ have the property that $P(\{c'^w_{\mu}\})\ket{\psi} \neq 0$ for any $\{c'^w_{\mu}\}$ with parity $\pi$. Since $\Pi_\mu C^w_{\mu}$ commutes with all link operators $K_j=X_{j,1}X_{j,2}$ (unlike the individual $C^w_{\mu}$), we can consider the eigenstates in sectors with simultaneously fixed eigenvalues of the link operators $\{K_j\}$ and $\pi$.

{\bf Proof}:\\
Let $\{K_j\}$ be the set of link operators which commute with every
$H(\{c^w_{\mu}\})$. Let $H(\{c^w_{\mu}\})\ket{\psi}=E \ket{\psi}$ where we can
thus assume that $K_j \ket{\psi}=\pm \ket{\psi}$. Consider $K_j
P\{(c^w_{\mu}\}) \ket{\psi}=P(\{c'^w_{\mu}\}) K_j \ket{\psi}=\pm
P(\{c'^w_{\mu}\}) \ket{\psi}$ where the eigenvalues $c'^w_{\mu}$ have changed
in sign on two adjacent white plaquettes compared to $c^w_{\mu}$. It is not
hard to see that there exists a product of link operators $K=K_{j_1}\ldots
K_{j_k}$ such that $K P(\{c^w_{\mu}\}) \ket{\psi}=\pm P(\{c'^w_{\mu}\})
\ket{\psi}$ for any pair of $\{c^w_{\mu}\}$ and $\{c'^w_{\mu}\}$ which have
the same overall parity, i.e., $\Pi_\mu c^w_{\mu}=\Pi_\mu c'^w_{\mu}$. This is because a link operator anticommutes with two plaquette operators $C^w_{\mu}$ and hence flips the eigenvalue for two $\mu$, furthermore one can generate any 
pattern of $\pm 1$ eigenvalues with fixed parity by applying multiple link operators. Since $K_j$ and $K$ are unitary, it implies that if some $P(\{c^w_{\mu}\})\ket{\psi}=0$, then all $P(\{c'^w_{\mu}\})\ket{\psi}= 0$ for $\Pi_\mu c'^w_{\mu}=\Pi_\mu c^w_{\mu}$ and thus $P_{\pi}\ket{\psi}=0$. Hence if $P_{\pi} \ket{\psi} \neq 0$, then all $P(\{c^w_{\mu}\})\ket{\psi}\neq 0$.

\clearpage
\newcounter{save}
\setcounter{save}{\value{NAT@ctr}}
\appendix


\begin{thebibliography}{10}

\bibitem{kitaev:top}
  A. Yu. Kitaev, Ann. Phys. (NY) {\bf 303}, 2 (2003), quant-ph/9707021.

\bibitem{wen:model}X.-G.~Wen, Phys. Rev. Lett. {\bf 90} (1), 016803 (2003). 

\bibitem{dennis+:top}
E. Dennis, A. Kitaev, A. Landahl, and J. Preskill,
 J. Math. Phys. {\bf 43}, 4452 (2002).

\bibitem{RHG:threshold}
R. Raussendorf, J. Harrington, and K. Goyal,
 New J. Phys. {\bf 9}, 199 (2007).

\bibitem{bombin_martindelgado}
H. Bombin and M. A. Martin-Delgado,
 J. Phys. A {\bf 42} (9), 095302 (2009).

\bibitem{fowler+:unisurf}
A. G. Fowler, A. M. Stephens, and P. Groszkowski,
 \pra {\bf 80} (5), 052312 (2009).

\bibitem{divincenzo_arch}
D. P. {DiVincenzo},
Phys. Scr. {\bf T137}, 014020 (2009).

\bibitem{kitaev:anyon_pert}
A. Yu. Kitaev,
 Ann. Phys. (NY) {\bf 321}, 2 (2006).

\bibitem{brell+:gadgets}
C. G. Brell, S. T. Flammia, S. D. Bartlett, and A. C. Doherty,
 New J. Phys. {\bf 13}, 053039 (2011).

\bibitem{kitaev:01}
A. Yu. Kitaev,
 Phys.-Usp. {\bf 44} (suppl.), 131 (2001),
 cond-mat/0010440.

\bibitem{fu:08}
L. Fu and C. L. Kane,
 Phys. Rev. Lett. {\bf 100}, 096407 (2008).

\bibitem{lutchyn:10}
R. M. Lutchyn, J. D. Sau, and S. Das Sarma,
 Phys. Rev. Lett. {\bf 105}, 077001 (2010).

\bibitem{oreg:10}
Y. Oreg, G. Refael, and F. von Oppen,
 Phys. Rev. Lett. {\bf 105}, 177002 (2010).

\bibitem{mourik:12}
  V. Mourik, K. Zuo, S. M. Frolov, S. R. Plissard, E. P. A. M. Bakkers, and L.
  P.  Kouwenhoven, Science {\bf 336}, 6084 (2012).

\bibitem{deng:12}
  M. T. Deng, C. L. Yu, G. Y. Huang, M. Larsson, P. Caroff, and H. Q. Xu,
   arXiv:1204.4130 (2012).

 \bibitem{rokhinson:12}
   L. P. Rokhinson, X. Liu, and J. K. Furdyna,
    arXiv:1204.4212 (2012).

\bibitem{hassler:10}
F. Hassler, A. R. Akhmerov, C.-Y. Hou, and C. W. J. Beenakker,
 New J. Phys. {\bf 12}, 125002 (2010); F. Hassler, A. R. Akhmerov, and C. W.
 J. Beenakker,
 \emph{ibid.} {\bf 13}, 095004 (2011)

\bibitem{sau:10}
J. D. Sau, R. M. Lutchyn, S. Tewari, and S. Das~Sarma,
 Phys. Rev. Lett. {\bf 104}, 040502 (2010).

\bibitem{beenakker:11}
C. W. J. Beenakker,
 arXiv:1112.1950 (2011).

\bibitem{alicea:11}
J. Alicea, Y. Oreg, G. Refael, F. von Oppen, and M. P. A. Fisher,
 Nature Phys. {\bf 7}, 412 (2011).

\bibitem{sau:11}
J. D. Sau, D. J. Clarke, and S. Tewari,
 Phys. Rev. B {\bf 84}, 094505 (2011).

\bibitem{heck:11}
B. van Heck, A. R. Akhmerov, F. Hassler, M. Burrello, and C. W. J. Beenakker,
New J. Phys. {\bf 14}, 035019 (2012).

\bibitem{halperin:11}
B. I. Halperin, Y. Oreg, A. Stern, G. Refael, J. Alicea, and F. von Oppen,
Phys. Rev. B {\bf 85}, 144501 (2012).

\bibitem{xu_fu:fracmodel}
C. {Xu} and L. {Fu},
 Phys. Rev. B {\bf 81} (13), 134435 (2010).

\bibitem{gladchenko}
S. Gladchenko {\em et al.}, Nature Physics {\bf 5}, 48--53 (2009).

\bibitem{starketal}
C.~Stark {\em et al.}, \prl {\bf 107}, 030504 (2011), J.~Wootton and J.~Pachos, \prl {\bf 107}, 030504 (2011).

\bibitem{FSS}
E.~Fradkin, M.~Srednicki and L.~Susskind, Phys.~Rev.~D {\bf 21}, 2885-2891 (1980).

\bibitem{familiar} The more familiar representation of the toric code is
  obtained by unitarily rotating $X \leftrightarrow Z$ on all qubits which are
  at positions $\nu\pm \hat{x}$ for, say, the gray plaquettes.

\bibitem{vsd:honey}
J. Vidal, K.P. Schmidt, S. Dusuel, Phys. Rev. B {\bf 78}, 245121 (2008).

\bibitem{book:ising}
A. D. B. K. Chakrabarti and P. Sen,
 {\em Quantum Ising Phases and Transitions in Transverse Ising Models\/}
 (Springer, 1996).

\bibitem{fu:10}
L. Fu, Phys. Rev. Lett. {\bf 104}, 056402 (2010).

\bibitem{BK:surface}
S. B. Bravyi and A. Yu. Kitaev,
 Quantum codes on a lattice with boundary, quant-ph/9811052 (1998).


\bibitem{toric} This is not true for the toric code in which plaquette
  operators are linearly dependent and a flipped sign of a link can lead to
  frustration.

\bibitem{koch:07}
J. Koch, T. M. Yu, J. Gambetta, A. A. Houck, D. I. Schuster, J. Majer, A.
  Blais, M. H. Devoret, S. M. Girvin, and R. J. Schoelkopf,
 Phys. Rev. A {\bf 76}, 042319 (2007).

\bibitem{bk:magicdistill}
S. Bravyi and A. Yu. Kitaev,
Phys. Rev. A  {\bf 71}, 022316 (2005).

\bibitem{loss:phys_honey}
F. L. Pedrocchi, S. Chesi, and D. Loss,
 Phys. Rev. B {\bf 84} (16), 165414 (2011).

\bibitem{lieb:flux}
E. H. Lieb,
 Phys. Rev. Lett. {\bf 73}, 2158 (1994).


\bibitem{proc} This procedure is similar to Kitaev's analysis of the
  honeycomb model \cite{kitaev:anyon_pert} where one obtains a noninteracting
  fermion model only if we consider an extended space of states, see also the
  discussion in \cite{loss:phys_honey}.

\bibitem{lieb} Perhaps one can prove this with techniques similar as in Ref. \cite{lieb:flux}. In this case we have a family of spin Hamiltonians (not fermionic) on a square-octagon lattice which depend on the parameters $c^w_{\mu}$ and one wishes to prove that the state of minimal energy is obtained when $\{c^w_{\mu}=+1\}$.

\end{thebibliography}
\end{document}